\documentclass[conference]{IEEEtran}
\usepackage{cite}
\usepackage{amsmath,amssymb,amsfonts}
\usepackage{algorithmic}
\usepackage{graphicx}
\usepackage{textcomp}
\usepackage{xcolor}
\usepackage[hyphens]{url}
\usepackage{fancyhdr}
\usepackage[bookmarks=true,breaklinks=true,letterpaper=true,colorlinks,citecolor=blue,linkcolor=blue,urlcolor=blue]{hyperref}
\usepackage{hyperref}
\usepackage{comment}
\usepackage{footmisc}
\usepackage{pifont}
\usepackage{makecell}
\usepackage{subcaption}
\usepackage{tabularx}
\usepackage{multirow}
\usepackage{rotating}
\usepackage{tikz}
\usepackage[inline]{enumitem}

\usepackage{hyperref} % for hyperlinking
\usepackage{lipsum} % for dummy text

\def\BibTeX{{\rm B\kern-.05em{\sc i\kern-.025em b}\kern-.08em
    T\kern-.1667em\lower.7ex\hbox{E}\kern-.125emX}}

\fancypagestyle{firstpage}{
  \fancyhf{}
  
  \fancyfoot[C]{\thepage}
}

\title{TDRAM: Tag-enhanced DRAM for Efficient Caching} 
\author{%
    \IEEEauthorblockN{%
        \begin{tabular}{c}
            Maryam Babaie\textsuperscript{1}, Ayaz Akram\textsuperscript{1}, Wendy Elsasser\textsuperscript{2}, Brent Haukness\textsuperscript{2}, Michael Miller\textsuperscript{2},\\
            Taeksang Song\textsuperscript{2}, Thomas Vogelsang\textsuperscript{2}, Steven Woo\textsuperscript{2}, Jason Lowe-Power\textsuperscript{1}\\
        \end{tabular}
    }
    \IEEEauthorblockA{%
        \begin{tabular}{c}
            \textsuperscript{1}Department of Computer Science, University of California, Davis \\
            \textsuperscript{2}Rambus Labs, Rambus Inc.
        \end{tabular}
    }
}
\begin{document}
\maketitle
\thispagestyle{firstpage}
\pagestyle{plain}

%%%%%% -- PAPER CONTENT STARTS-- %%%%%%%%

\begin{abstract}

  As SRAM-based caches are hitting a scaling wall, manufacturers are integrating DRAM-based caches into system designs to continue increasing cache sizes.
  While DRAM caches can improve the performance of memory systems, existing DRAM cache designs suffer from high miss penalties, wasted data movement, and interference between misses and demand requests.
  In this paper, we propose TDRAM, a novel DRAM microarchitecture tailored for caching.
  TDRAM enhances HBM3 by adding a set of small low-latency mats to store tags and metadata on the same die as the data mats.
  These mats enable fast parallel tag and data access, on-DRAM-die tag comparison, and conditional data response based on comparison result (reducing wasted data transfers) akin to SRAM caches mechanism.
  TDRAM further optimizes the hit and miss latencies by performing opportunistic early tag probing.
  Moreover, TDRAM introduces a flush buffer to store conflicting dirty data on write misses, eliminating turnaround delays on data bus.
  We evaluate TDRAM using a full-system simulator and a set of HPC workloads with large memory footprints showing TDRAM provides at least 2.6$\times$ faster tag check, 1.2$\times$ speedup, and 21\% less energy consumption, compared to the state-of-the-art commercial and research designs.
  
\end{abstract}
% License information
\begin{figure}[b]
    \hrule
    \vspace{0.5em}
    \raggedright
    \small This work is licensed under the Creative Commons Attribution 4.0 International \href{https://creativecommons.org/licenses/by-sa/4.0/}{(CC-BY-SA 4.0)} license.
    Authors reserve their rights to disseminate the work on their personal and corporate websites with the appropriate attribution.
\end{figure}

\section{Introduction}
\label{intro}

Today's computers, equipped with significant processing capabilities and memory capacities, aim to fulfill the requirements of high-performance computing (HPC) tasks such as machine learning and artificial intelligence.
To find the right balance between performance and capacity, manufacturers have embraced heterogeneous memory systems.
These systems combine high-performance memories like HBM with high-capacity memories with lower performance.
The introduction of interconnect technologies like Compute Express Link (CXL) is increasing memory heterogeneity with local and remote memory pools.
Intel's Sapphire Rapids CPU demonstrates this strategy by employing on-package HBM DRAMs as cache~\cite{sharma2023system,sapphire}.
This approach effectively boosts cache capacities and tackles the scaling limitations encountered by SRAM~\cite{deathOfSRAM}.
The expanded cache capacity and enhanced bandwidth provided by HBM DRAMs offer the potential for improved data locality, without programmers intervention.

However, the potential benefits of DRAM caches have not borne fruit.
Previous studies of DRAM caches have shown that using standard DRAM devices as a cache can slow down applications with large memory footprints and high miss rates~\cite{hildebrand2021case, raybuck2021hemem}.
The current designs of DRAM caches, such as those in Intel's Cascade Lake DRAM cache~\cite{cascadeLake, hildebrand2021case}, store tag and metadata together with the cache line  data in the same DRAM.
Storing tags and data together reduces hit time for read demands~\cite{qureshi2012fundamental}, but significantly increases miss penalties since a separate DRAM read is necessary to retrieve tag and metadata for hit/miss and status information, causing contention with demand reads.
Also, \emph{all write requests, including those hitting on the cache, require a DRAM read} to fetch tag and data to ensure dirty data is not overwritten, further exacerbating the contention and causing expensive turnaround bubbles on the data bus~\cite{arafra2019cascade}.
These extra accesses for read-misses and write demands increase: (i) latency for demand misses, (ii) contention in the read buffer which extends the queue occupancy time, and (iii) wasted data movement and energy consumption.

Today's applications that require large capacity memories have high miss rates in DRAM caches which cause these issues to significantly affect the workload's performance.
SRAM caches cannot scale to the capacities required by today's applications, and thus it becomes imperative to enhance existing DRAM cache designs to address these challenges.

In this work, we introduce TDRAM (\textbf{T}ag-enhanced \textbf{DRAM}), a DRAM microarchitecture specifically tailored for caching purposes.\footnote{In the recent past, it was uneconomical for DRAM manufacturers to modify the core DRAM microarchitecture. However, specialty DRAMs are becoming increasingly common (e.g., Samsung's Aquabolt~\cite{samsungAquabolt}, Micron's Automata Processor~\cite{want2016ap} and RLDRAM~\cite{rldram3}).}
TDRAM enhances HBM3 by adding a set of small low-latency mats to store tags and metadata on the same die as the data mats.
These mats enable faster access than the data mats through the reduction of wordline and bitline lengths.
The additional on-die storage is sufficiently large to accommodate the tag and metadata for all DRAM cache lines; thus, the tag store scales with the data capacity.
By placing the tags in separate mats on-die, TDRAM enables rapid on-die tag checking which reduces the latency for demand misses, mitigates contention on the DRAM read queue, and decreases wasted data transfers (and thus energy).

TDRAM extends HBM3's interface in three ways:
(1) It adds a unidirectional hit-miss (HM) bus to its interface to transfer the tag check result and metadata to the controller.
(2) It adds two new commands to the HBM3's protocol: \textit{ActRd} and \textit{ActWr}, which access both tag and data mats in lockstep.
These commands check the tag for the block and only send data to the controller when it is needed.
(3) We add a \textit{flush buffer} to store conflicting dirty data from write misses which eliminates costly turnaround delays on the data bus and immediate cache line data transfer to the controller for write requests.
The overhead of this new design compared to HBM3 is 192 pins (out of 1,972 existing pins, a 10\% increase) and 8.24\% total die area.

TDRAM further improves performance by implementing \textit{early tag probing}, which opportunistically performs tag checks (without data access) in otherwise unused command and HM bus slots.
Tag probing returns early hit-miss and status indication of a demand access, allowing certain operations (e.g., main memory access for read demand misses) to start earlier.
Early tag probing also reduces request queue occupancy time by removing misses from the queue early, allowing other demands to proceed with fewer stalls.

TDRAM is orthogonal to many prior works that focus on improving DRAM caching performance by adding predictors, prefetchers, tag caches, modifying coherence protocols, bypass policies, and other application-specific mechanisms~\cite{yang2016etag, hameed2020improving, chou2015bear, behnam2022adaptively, behnam2018r}.
TDRAM is designed in a way that all of these techniques can be applied on top of it to further improve its performance.
Overall, TDRAM enables a perfectly scalable HBM-based cache with a cohesive caching paradigm akin to processors' SRAM-based caches.
Thus, TDRAM focuses on optimizing the core and fundamental DRAM cache operations by eliminating inefficiencies in existing designs. %~\cite{hildebrand2021case}.

We have extensively modeled TDRAM in the gem5 simulator\cite{lowepower2020gem5}, for a detailed full-system cycle-level timing analysis.
Our evaluations using scientific and graph analytics applications with large memory footprints, have shown TDRAM provides 2.6\texttimes~faster tag check and 1.2\texttimes~speedup and at least 21\% less energy consumption, compared to the commercial and research designs such as Intel's Cascade Lake and Alloy.

In this paper, we make the following contributions:

\vspace{1em}
\textbf{Microarchitecture}
\begin{itemize}[leftmargin=0.3cm]
\item We propose a new \textit{HBM3}-based DRAM microarchitecture, TDRAM, designed for caching with on-DRAM-die tag management, to enable \textit{perfectly scalable} DRAM caching where the tag storage scales with data capacity.
\item We extend the HBM3 interface with a unidirectional \textit{Hit-Miss bus} to transfer tag check results and metadata from DRAM to the controller, decoupling them from data transfer.
\item We add a \textit{flush buffer} to hold conflicting dirty data from write misses which eliminates both costly turnaround delays on the data bus and immediate cache line data transfer to the controller for write requests.
TDRAM opportunistically sends them to the controller when data bus is idle or in read-state.
\end{itemize}

\textbf{Protocol}
\begin{itemize}[leftmargin=0.3cm]
\item We integrate two new commands to HBM3's protocol, \textit{ActRd} and \textit{ActWr}, which enable parallel independent access to both tag and data banks.
      The protocol \textit{selectively streamlines data} to the controller only when necessary based on tag comparison, reducing bandwidth bloat.
\item We propose \textit{opportunistic early tag check} mechanism in unused HM and command bus slots, to minimize queueing delay for tag check, reducing buffer contention and optimizing miss latencies.
      This mechanisms does not access data banks.
      If this tag check results in miss, then TDRAM initiates backing store access immediately, if necessary; and avoids future cache line data access, if not necessary.
\end{itemize}

\textbf{Evaluation}
\begin{itemize}[leftmargin=0.3cm]
\item We demonstrate DRAM caching using existing designs cause \emph{slowdown} while TDRAM provides 1.11\texttimes~speedup.
\item We show TDRAM reduces energy consumption by 21\%, since its conditional data response to the controller removes wasted data transfers.
\item We analyze performance of DRAM caching in disaggregated systems with remote main memories and show TDRAM provides 1.14\texttimes~speedup compared to the state-of-the-art commercial DRAM cache, for low miss ratio applications.
\end{itemize}

\begin{table*}[t]
      \caption{Comparison of TDRAM with related work. Tag storage can be on the processor die (e.g., in SRAM or eDRAM) or off die (e.g., on DDR or HBM device). The tag check can occur before the off-die memory controller, within the memory controller, or in the off-chip data storage. The area overhead on the processor die depends on the tag storage location. Some designs require extra hardware (e.g., tag cache, prefetchers) or significant changes to the coherence protocol (Extra hardware row). The tag capacity can either scale with the amount of data stored in the cache or not. Some designs require multiple (tag and/or off-chip) device accesses to complete read and write hits. TDRAM has low are overhead, no extra hardware, scales tag capacity with data capacity, and requires only one device access for read and write hits.}
      \vspace{-0.5em}
      \centering
      \includegraphics[width=\linewidth]{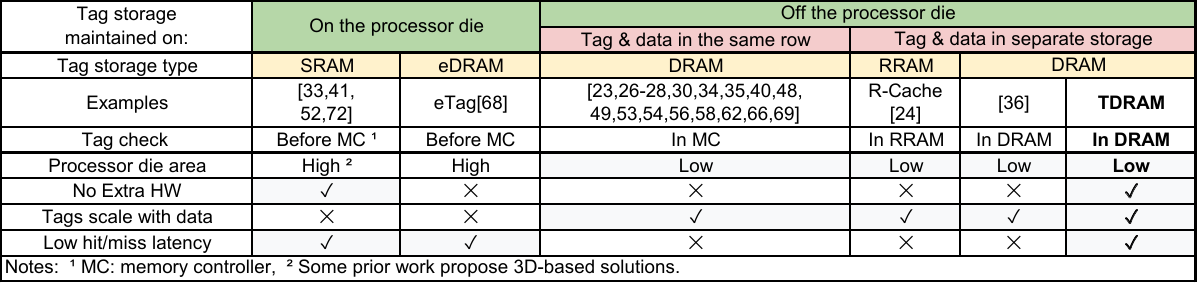}
      \vspace{-2em}
      \label{tab:related}
\end{table*}

\section{Background and Motivation}
\label{sec:motivation}

\subsection{HBM3 Architecture}
\label{sec:hbm3desc}
This section provides an overview of HBM DRAMs as the basis of TDRAM.
HBM provides the highest bandwidth and capacity of any single DRAM package in mass production.
HBM3 DRAMs stack multiple DRAM die into a single package, and can support up to 64~GiB capacity using 12 to 16-high stacks~\cite{isscc16High}.
These devices provide up to 1024~GiB/s of bandwidth when running at 8~Gbps across 16 independent channels with 64b data (DQ) and 10b Row command (R) and 8b Column command (C) buses.
Each DQ channel can be split into two 32-bit pseudo-channels (PCs) that share the same R and C buses, with each PC providing 32B access granularity~\cite{hbm3jedec,rambushbm3,samsunghbm3,synopsyshbm3}.
Each channel includes 38 additional signals for clocks, strobes, ECC, redundancy, and other functions.
The wires connecting the high pin count interface between the DRAM and host (1024 DQs, 288 command/address (CA) buses, and more than 650 pins for additional channel and global functions), are implemented in technologies such as TSMC's InFO or silicon (e.g., a silicon interposer) to support the high pin and trace densities required for this technology.

HBM DRAMs are hierarchically organized, storing data bits in arrays of capacitors.
DRAM bit cells are grouped into $rows$ or $pages$, with multiple $rows$ aggregated into $mats$, and $mats$ organized into 2D structures called $banks$.
Decoded row and column addresses identify bits within a bank.
Multiple banks form a $bank$ $group$ that share some resources.
Back-to-back accesses to the same bank group require longer latencies to allow these shared resources to free up, while accesses to different bank groups enable lower latencies.

A command decoder receives commands and addresses from the memory controller over a CA bus.
When data is read from a DRAM, an activate command provides a row address to move all bits in one row of a bank to sense amplifiers (or sense amps).
A separate read command provides a column address to select a subset of the bits from the sense amps to be returned across the DQ bus.
Write commands work similarly, providing data to be written into the DRAM.

\subsection{Tag Management in Existing DRAM Caches}
\label{sec:tagMetaStorMngmt}
Numerous studies have investigated the management of tag and metadata (referred to collectively as \textit{tag}) in hardware-managed DRAM caches and Table~\ref{tab:related} compares some of these prior works to TDRAM.
Also, DRAM cache products, like Intel's Xeon series, offering gigabytes of DRAM cache, are available in the market.
In terms of storage size, a 64~GiB block-based DRAM cache requires 3~GiB of storage for 3B tag per 64B blocks. 
This is far beyond the cache sizes in high-end CPUs by AMD (384~MiB in EPYC 9654P~\cite{EPYCcache}) and Intel (105~MiB in Xeon Platinum 8468H~\cite{intelcache}) today.
While SRAM caches are hitting a scaling wall, tags-in-SRAM solutions (e.g., on processor die) will add to the area overhead and price~\cite{zhao2007exploring, inoue20113d, madan2009optimizing, ghosh2007smart}.
Moreover, solutions that put tags on the processor die, e.g., eTag~\cite{yang2016etag}, severly limits scalability of DRAM cache capacity by tying it to the tag capacity that processor chip can provide.
Where Aurora supercomputer~\cite{aurora} offers over 64~GiB HBM per CPU, moving towards highly-scalable HBM-based DRAM caches is the direction of future.

Previous studies suggest storing tags in the same cache line that data resides~\cite{behnam2022adaptively, bojnordi2019retagger, meswani2015heterogeneous, yin2015cooperatively, loh2012challenges, sim2013resilient, sun2009design, park2012efficient, hameed2013simultaneously, loh2012supporting, chou2015bear, meza2012enabling, qureshi2012fundamental, chou2014cameo, gulur2014bi, el2013dual, huang2014atcache}.
For instance, in Alloy cache instead of accessing 64B block, 72B (plus 8B ignored) must be accessed, causing misalignment in column layout within DRAM rows and leaving unused bits that causes scalability overhead.
In Intel's Xeon Series (e.g., Cascade Lake), tags are stored in the unused bits of ECC in commodity DRAM devices~\cite{hildebrand2021case}.
However, ECC bits are not designed for this purpose. 
These designs depend on a DRAM read to access tag which can create a serialization of tag and data access (e.g., in write-hits), increasing bandwidth bloat.

Some prior work proposed to store tags in separate storage on DRAM die.
R-Cache uses resistive RAM for tag storage~\cite{behnam2018r}.
Since tag access is on the critical path (i.e., data access in DRAM cache depends on the tag comparison result), the tag read and update latencies must be minimized.
Resistive RAM cannot provide the required speed.
Moreover, tag and metadata are subject to frequent updates, which can wear out resistive RAM quickly.
Other works suggested DRAM-based tag storage~\cite{hameed2020improving,yang2016etag}.
These works optimize tag management and data layout in DRAM rows for associative caches which require multiple tag comparisons, and both activate tag and data regions in parallel.
However, they have to delay the start of column operations till tag comparison logic finds the corresponding column, which in fact internally ties the data access to the tag access.
Most importantly, they fall short in providing an efficient mechanism to handle write misses to dirty cache lines that necessitates a data read before write for correctness.
As a result, these solutions have to rely on either speculative mechanisms (e.g., predictors and DRAM bypassing unit with application-specific hardcoded designs~\cite{hameed2020improving}), or require deep cache coherence protocol changes (e.g., for the LLC to send clean writeback messages to the memory controller)~\cite{lowe2017heterogeneous}.
Changes to the cache coherence protocol ties the designs of coherence protocol, the memory controller, and the DRAM, which we avoid in this work.

\subsection{Opportunities to Improve DRAM Cache Designs}
\label{sec:hitMissPenalty}
DRAM cache hit/miss latencies vary based on system design, influenced by factors like tag storage and access mechanisms. 
Tag check latency is always on the critical path of servicing a memory demand. 
The prior designs storing tag in the DRAM cache line~\cite{qureshi2012fundamental, jevdjic2014unison} or in the ECC bits of DRAM (e.g., Intel's Cascade Lake), must perform a DRAM read that retrieves the cache line's tag and data, simultaneously. 
These designs aimed for parallelizing the tag and data access to improve hit latency.
Through some experiments we show their inefficiencies.

\begin{figure}
  \centering
  \includegraphics[width=\linewidth, scale=0.8]{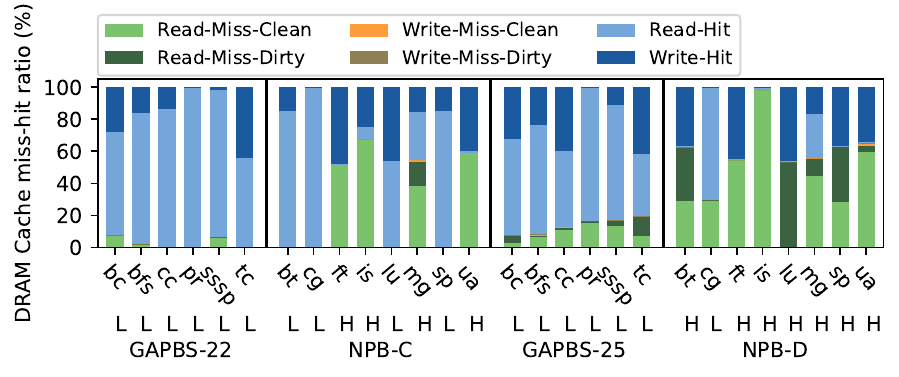}
  \vspace{-2em}
  \caption{The breakdown of hit and miss ratios of DRAM cache. 
  The letters show high or low miss ratio.}
  \label{fig:hmBreak}
  \vspace{-1.5em}
\end{figure}

For the experiments we consider Intel's Xeon Max series~\cite{sapphire}, rounded up to 64 cores and 64~GiB of HBM (1~GiB per core, in DRAM cache mode). 
Using gem5 simulator~\cite{lowepower2020gem5}, we modeled $\frac{1}{8}$ of the target system. % and a DRAM cache architecture similar to the prior designs that stores tag in DRAM. 
\textit{In this work, for the baseline to represent existing designs storing tags in DRAM, we choose Intel Xeon series DRAM cache (e.g., Cascade Lake), recognized as the state-of-the-art real commercial product in this domain which implements a direct-mapped insert-on-miss cache~\cite{hildebrand2021case}.}
We executed 28 HPC multithreaded workloads from GAPBS~\cite{beamer2015gap} and NPB~\cite{bailey1991parallel}.
These workloads' memory footprints are 0.1--80~GiB, while the total DRAM cache size remains at 8~GiB.
We conducted full-system simulation and we employed LoopPoint technique~\cite{sabu2022looppoint} for precise control over workloads execution phases. 
Notably, our experimental setup differs from previous works, allowing us to uncover pathological pathways not observed in prior studies.
\S\ref{sec:method} provides more details about our methodology.

\subsubsection{\textbf{DRAM Cache's Increased Hit Latency}}
\label{sec:motivHitLat}
In DRAM caches that use standard DRAM devices with tags stored in the device, the cache hit latency almost equals the DRAM read latency for LLC read misses.
For LLC writebacks, i.e., evicting dirty data from LLC, the hit latency involves a DRAM read latency (to retrieve the tag) and then a DRAM write is issued (for writing incoming data into the cache). 
Previous efforts aimed to parallelize tag and data accesses for each memory request~\cite{qureshi2012fundamental}.
However, this parallelization is compromised for all LLC writebacks, even those hitting on the DRAM cache.
The reason is the controller must read the tag (in which the data is also read) for tag comparison and it will not issue the incoming data write into the cache until after DRAM read is ready in the controller and tag check is done.
Note that the data read happens because in the commodity DRAMs the only mechanism to fetch tag is to read the whole cache line data, regardless of incoming write data size.
In this case, the DRAM read remains on the critical path of write demand and causes access amplification (bandwidth bloat).
Prior work reported DRAM caches' access amplification can reach to as high as 5 accesses~\cite{hildebrand2021case}.
Figure~\ref{fig:hmBreak} illustrates the miss ratio percentage of the DRAM cache and its breakdown in our experiments.
As shown in dark blue color, in the majority of workloads, the number of LLC writebacks hitting on the DRAM cache is significant, potentially affecting the hit latency of the DRAM cache.

\begin{figure}
  \centering
  \includegraphics[width=\linewidth, scale=0.8]{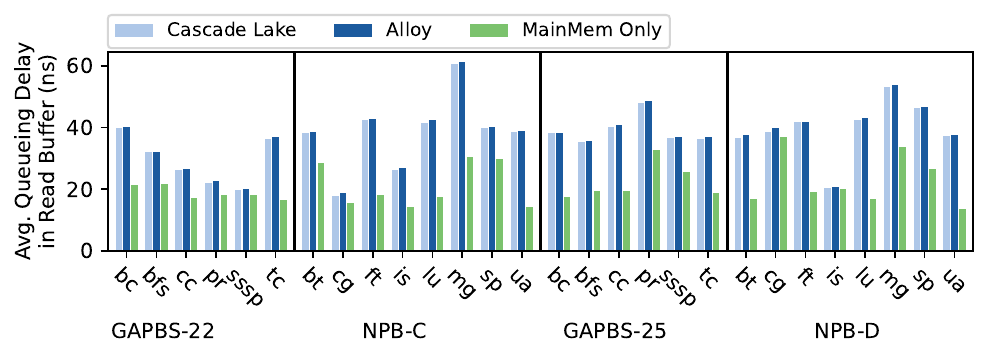}
  \vspace{-2em}
  \caption{The average queueing delay of read demands in the read buffer of controller, 
          in Intel's Cascade Lake and Alloy DRAM caches,
          compared to the system having a main memory only (no DRAM cache).
          This time marks the waiting time requests spend in the buffer before accessing the memory.}
  \label{fig:queueDel}
  \vspace{-1.5em}
\end{figure}

\subsubsection{\textbf{DRAM Cache's Increased Miss Latency}}
The process time of a memory demand is sum of two components: $Queueing Delay + Memory Access Latency$. 
Memory access latency is the time it takes since a read/write command is issued for a demand, until the data is available on the DQ bus.
The queueing delay marks the waiting time requests spend in a buffer before a read/write command is issued.
In current DRAM caches, all read and write requests (i.e., LLC's read misses and writebacks) must undergo a DRAM read to fetch the tag.
The controller handles these DRAM reads, including those for LLC writebacks, in the same read buffer.
This arrangement heightens contention in the buffer, increasing queueing delay of all demands.

Figure~\ref{fig:queueDel} displays the average queueing delay of all DRAM reads in state-of-the-art DRAM caches, compared to a system solely equipped with main memory (no DRAM cache).
As depicted, the bars are significantly higher in the DRAM cache system compared to system relying solely on main memory.
The main reason is that every read and write demand has to start by reading a tag in DRAM cache, which increases contention in the read buffer and bank conflicts when the tag reads occur.

This extended delay directly impacts the tag check latency for all read demands that miss in the DRAM cache, leading to a delay in fetching the missing line from the main memory where the response to LLC resides.
In simpler terms, it extends the miss latency of the DRAM cache.
This latency is crucial for LLC read misses, as their processing time in the DRAM cache contributes to the miss penalty of LLC, which is observed by the CPU.
This directly influences the overall system throughput.
As Figure~\ref{fig:hmBreak} shows, the number of read misses (in dark/light green) in the DRAM cache is significant.

\subsubsection{\textbf{Increased Bandwidth Bloat and Energy Consumption}}
\label{sec:motivBWbloat}

\begin{figure}
  \centering
  \includegraphics[width=\linewidth, scale=0.8]{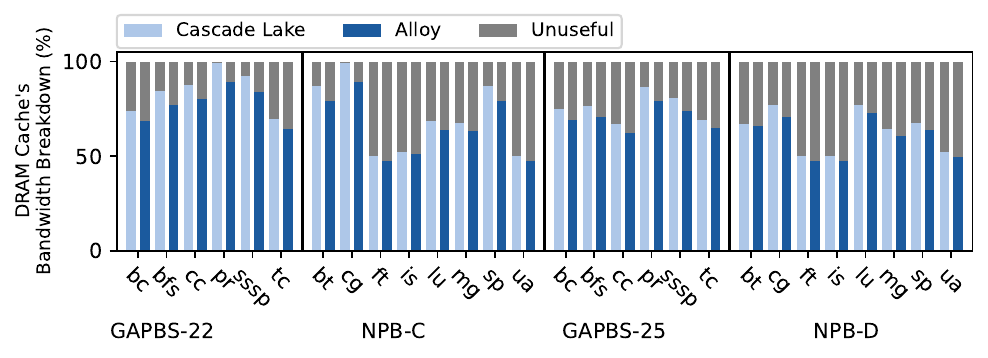}
  \vspace{-2em}
  \caption{Intel's Cascade Lake commericial and Alloy DRAM caches bandwidth, broken to useful and unuseful data movement, normalized to total system bandwidth.
           In all read/write misses to a clean line and write hits, after tag comparison (which also retreives data) the controller immediately discards the data (serving no purpose), shown as unuseful.
           Alloy has a longer burst length than Cascade Lake, which increases the unuseful data movement.}
  \label{fig:extraDataTransfer}
  \vspace{-1.5em}
\end{figure}

The read data during tag access in the DRAM cache is only beneficial for read demands hitting the cache or for demands missing to a dirty cache line.
\textit{In cases of read/write misses to a clean line and write hits, the controller discards the read data immediately after tag comparison, serving no purpose.} 
In such cases, existing DRAM cache designs introduce data movement overheads by:
(i) keeping DRAM's command bus and banks busy, and 
(ii) occupying data bus for unnecessary data transfers.
These extra communications between the DRAM and the controller result in bandwidth bloat\cite{chou2015bear}, wasting energy.
This waste exacerbates as the miss ratio rises.
Figure~\ref{fig:extraDataTransfer} quantifies the relative amount of wasted data movement during the tag check process.
Notably, in many applications (e.g., $ft, is, mg, ua$) the wasted data movement is significant.

Moreover, in cases where the read data is a dirty line, it is unnecessarily part of the critical path of servicing a demand.
A thoughtful design could put such accesses off the critical path while ensuring correctness.
Figure~\ref{fig:hmBreak} illustrates that the memory demands not using the read data in tag access (i.e., write-hits, read-miss-cleans, write-miss-cleans) are common.
Notably, write demands that miss to a dirty line in DRAM cache are very rare, indicating an opportunity to eliminate data reads in tag checks on write demands. 

From this preliminary analysis, we have the following goals when constructing a cache-optimized DRAM architecture:

\begin{enumerate} [leftmargin=0.4cm]
  \item \textit{Reduce the hit latency} of DRAM cache by optimizing the tag check mechanism and write-hits;
  \item \textit{Reduce the miss latency} of DRAM cache specifically for reads by reducing tag check and queueing delays;
  \item \textit{Reduce the wasted data movement} on write-hits, read-miss-cleans, and write-miss-cleans to save energy; and
  \item Support write-miss-dirty (i.e., we cannot simply overwrite on writes) for correctness not necessarily performance since they are uncommon.
\end{enumerate}
\vspace{-0.2em}

\section{Tag-enhanced DRAM Design}
\label{sec:design}

\begin{figure*}
  \centering
  \includegraphics[width=\linewidth]{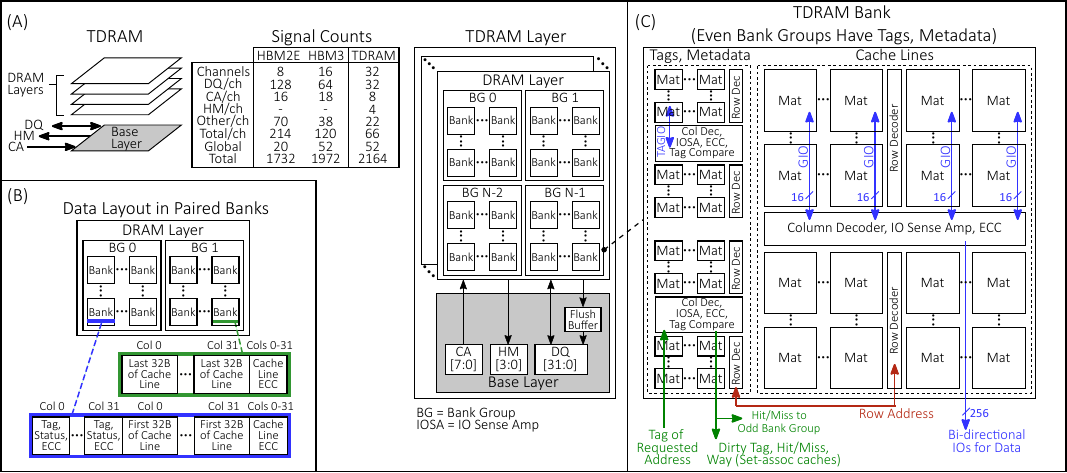}
  \caption{TDRAM's architecture and bank organization.}
  \label{fig:bankOrg}
\end{figure*}

In this section, we describe the microarchitecture of TDRAM, a new DRAM crafted to fulfill the requirements of DRAM caching in contemporary server memory hierarchies.
TDRAM is designed in the same vein as other custom DRAMs such as Samsung Aquabolt~\cite{samsungAquabolt}, Micron Automata Processor~\cite{want2016ap}, and RLDRAM~\cite{rldram3}.
Given the slowdown in SRAM scaling, industry is already integrating DRAMs as caches~\cite{sapphire,arafra2019cascade,knl:Sodani:2016} and announced future DRAM devices specialized for caching~\cite{samsungllcdram}.
TDRAM is a novel microarchitecture for specialized DRAM devices in this track.

TDRAM adheres to three principles:
(i) reducing hit and miss latencies,
(ii) minimizing bandwidth bloat and communications between cache and controller to save energy, and
(iii) easily scalable and providing high bandwidth.
TDRAM is based on HBM3 DRAM architecture (see \S\ref{sec:hbm3desc}).

The following sections describe the microarchitectural details of TDRAM. 
First, we explain the interface of TDRAM that connects the device and the controller. 
Second, we elaborate on the internal structure of TDRAM that provides the additional on-die tag storage and fast tag check mechanism while maintaining the data access per usual HBM3. 
Third, we demonstrate the protocol of TDRAM that supports new combined commands to access both tag and data banks in lockstep (true parallelization), providing conditional data response based on the tag comparison result. 
Finally, we explain the opportunistic behaviors of TDRAM on: (i) early tag probing, and (ii) sending evicted dirty data on write misses to the controller when data bus is idle.

\subsection{TDRAM's Interface}
\label{sec:interface}
TDRAM leverages the HBM3 interface and introduces three changes as shown in Figure~\ref{fig:bankOrg}A: 
(i) the R and C buses are merged into a single CA bus (i.e., like DDR DRAMs), 
(ii) each of the 32 PCs is converted to an independent channel with its own 8b CA bus and 32b DQ bus, and 
(iii) a 4b Hit-Miss (HM) bus is added to each channel.
Converting PCs to independent channels simplifies memory controller design, as each PC already has its own memory controller~\cite{rambusHBM3MC} and command/address arbitration for the shared R and C buses in HBM3 can be removed.
Each of the 32 channels has a 32b DQ bus and an 8b CA bus over which both row commands and column commands are sent.
Data transfers are protected by ECC and redundancy as is done in HBM3.

The CA bus runs at the same speed as the DQ bus and the protocol is designed to leave enough CA bus bandwidth available for additional operations like early tag probing (\S\ref{sec:earlyTagProbe}).
The 4b HM bus is unidirectional, runs at the same speed as the DQ bus, and TDRAM uses it to communicate back to the host: the results of on-die tag comparisons (hit/miss), status information (valid/invalid, dirty/clean, etc), and tag of dirty data to be written back to the main memory.
A 64~GiB direct-mapped DRAM cache can support 1 petabyte address space using a 14-bit tag.
The HM bus runs at full data rate and the data packets are much longer than the HM bus occupancy for a single transaction.
Thus, the tags and metadata can be transferred over HM bus in a number of beats without BW becoming an issue.
Each channel has 22 additional signals (clocks, strobes, ECC, etc.) and the DRAM has 52 additional global signals (reset, IEEE1500, etc.) for a total of 2164 signals, an increase of 9.7\% over HBM3.
The Signal Counts table in Figure~\ref{fig:bankOrg}A shows the details of TDRAM's signals overhead compared to HBM3.
The HBM3 package has 320 unused bump sites in the area for address and data signals~\cite{hbm3jedec}, more than enough to accommodate the additional 192 signals (2b CA + 4b HM = 6, per 32-bit channel) in TDRAM, allowing TDRAM to use a similar package.

\subsection{TDRAM's Internal Architecture}

\subsubsection{\textbf{Data Storage and Access Granularity}}

TDRAM maintains the standard bank microarchitecture found in HBM3 devices for data storage.
Since server CPUs from Intel and AMD operate on 64B cache-lines, but HBM DRAMs are designed to provide 32B granularity, TDRAM pairs banks in different bank groups and staggers accesses to them to achieve 64B granularity at lower latencies.
Figure~\ref{fig:bankOrg}B shows the layout of these paired banks.
The controller views the paired banks as one logical bank and schedules accesses accordingly.
To simplify the management of paired banks, the controller issues a single command (e.g., activate, read, etc.) and the logic on the base die replicates it, staggering it in time across the bank pair.
Pairing banks across bank groups in this way simplifies the controller management since the design eliminates the back-to-back accesses to the same bank group.

\subsubsection{\textbf{On-Die Tag Storage}}
\label{sec:tagBanks}

TDRAM stores tags and metadata (refered collectively as tag), and their ECC in a separate structure on the same die as the cache line data. 
TDRAM uses a set of small low-latency mats to store tags to provide fast access. 
The size of tag storage is much smaller than the size of data storage (about $\frac{3}{64}$, i.e., 3B tag per each 64B cache line).
The smaller size allows these mats to have shorter wordline and bitline lengths (than the data mats), similar to the design of Reduced Latency DRAM (RLDRAM)~\cite{rldram3}. 
Latency improvements with scaled mats are discussed in~\cite{son2013reducing}.
Our design scales the tag mats by $\frac{1}{2}$ in each direction, reducing the wordline delay time and bitline charge sharing completion time.
Centralized decodoer and IOSA structures further improve the latency.
These low-latency mats allow parallel tag and data lookup, with hit/miss being determined before the data becomes available in the data mats.
The special mats are placed at the edge of each bank (Figure~\ref{fig:bankOrg}C).

As an alternative design, the reduced latency tag arrays can also be implemented on a separate die within the TDRAM stack.
However, an advantage of placing tags on the same die as the cache-line data is that tag storage scales with data storage.
For the remainder of this paper we assume the tags are on the same die as the data.

\subsubsection{\textbf{Metadata Access and Tag Comparison}}
We add two new DRAM commands to the HBM3 command set: \textit{activate-read (ActRd)} and \textit{activate-write (ActWr)}.
When a ActRd or ActWr command is issued to a bank, the tag mats are activated in parallel with the data mats.
To avoid sending tags and metadata back to the controller, TDRAM uses on-die tag comparators implemented in the IOSA area of the tag mats to determine whether an access is a hit or a miss.
Then, the HM result is routed to the periphery of the chip for output on the HM pins.
The HM result is also sent to the column decoders of the data mats where it is used to gate the column decode logic using the same hardware as SALP~\cite{kim2012salp}.
If the tag comparison results in a hit or in a miss to a dirty cache-line for read demands, the data is transferred through the DQ bus.
If the tag comparison results in a miss to a clean cache-line, the column decode does not happen and no data is transferred on the DQ bus, saving energy.
Furthermore, to improve reliability, there are ECC bits for the tag which are analyzed and corrected if needed by on-DRAM-die circuitry as in the baseline HBM3~\cite{hbm3jedec}.
Figures~\ref{fig:readHit}, \ref{fig:pipelinedReadsTrans}, \ref{fig:wrDirty}, and~\ref{fig:tagProbTrans} show the timing transactions of read and write operations in TDRAM and are discussed in detail in \S\ref{sec:protocol}.

\subsubsection{\textbf{Direct-Mapped \& Set-Associative TDRAM}}
\label{sec:setAssocTDRAM}
The architecture of TDRAM applies equally well to direct-mapped and set-associative caches.
On-die tag comparison can provide greater benefits for set-associative caches.
In Figure~\ref{fig:bankOrg}, if pairs of bank groups (e.g., 0 and 1, 2 and 3, etc) form two ways of a set, tag comparisons can be performed in parallel if each way has its own comparator.
A signal from the matching way is sent to the internal control logic to select the proper column in the data mats.
Implementations without on-die tag comparators send all tags in the set back to the controller, incurring additional latency and energy, and the controller subsequently sends a request for the proper column to the DRAM, again incurring additional latency and energy consumption~\cite{loh2012supporting}.

\subsubsection{\textbf{Tag Mats Timing Values}}
\label{sec:setTagTimings}
TDRAM architecture minimizes the tags access latencies using small low-latency mats as discussed in \S\ref{sec:tagBanks}.
\textit{In our evaluation, we use timing parameters for the tag mats that are loosely based on RLDRAM technology.}
We choose to base our timings on public datasheets as the timing parameters for RLDRAM are close to the \textit{proprietary analysis we conducted}.
Through discussions with DRAM designers, we validated the internal timing values.
Table~\ref{tab:conftable} shows a list of timing values we used.
The RLDRAM spec values (e.g., tRL=15ns and tRC=8ns) match, or are more optimistic than, our values (e.g., tRCD\_TAG+tHM=15ns and tRC\_TAG=12ns).
Furthermore, these values and internal TDRAM timings were also correlated with prior work analyzing the use of smaller mats~\cite{son2013reducing}. 

Additionally, tHM\_int = tCCD\_L+tHM\_detect (which is a fast equal comparison).
Address comparisons are already done in DRAMs today to quickly determine if every row or column address that the DRAM receives is a repaired row or column.
We set tHM\_detect to 0.5ns (one 2GHz clock cycle) for the fast equal comparison based on discussions with expert DRAM designers.
The use of tHM\_int depends on tRCD since a read operation cannot occur until tRCD is met.
In our design, tRCD = 12ns which is longer than tRCD\_TAG+tHM\_int=10ns, effectively hiding the tag access and hit/miss detection latency.
tHM\_int was also correlated with prior work~\cite{son2013reducing}, which breaks ACT-to-data delay into: 47\%-sensing, 26\%-address-decode, 20\%-MUXing (transfer+rate-conversion), and 7\%-IO. 
The column decode is done in parallel to sensing (tRCD) with our ActRd/ActWr commands.
Finally, the I/O delay is not relevant to internal timing.
Only a portion of the MUXing delay, the delay to move data out of the IOSA, is relevant to internal timing, and this delay is optimized with smaller mats to achieve tHM\_int=2.5ns, including HM detect logic.

To ensure dirty data is not overwritten in the SA, tRL\_CORE (used in write operations as illustrated in Figure~\ref{fig:wrDirty}) needs to be less than or equal to intRD-to-WR\_data\_Delay+tBURST/2=9ns.
We performed our evaluation using tRL\_CORE=tCCD\_L=2ns.

\subsubsection{\textbf{Tag Storage Area Overhead}}
\label{sec:areaOverhead}
A 64~GiB direct-mapped DRAM cache can support 1 petabyte address space using a 14-bit tag. We assume 3B of tag and metadata for each 64B cache line.
Tags are stored only in one bank group of the pair (the even-numbered bank groups), and the result of the tag comparison is communicated to the other (odd-numbered) bank group through an internal bus. 

We estimate the die size impact of tag storage, control logic, and on-die comparison as follows. 
HBM3 DRAMs store an additional 6B of information (2B metadata and 4B parity) for every 32B of data (i.e., total column size is 38B) across 19 mats as shown by Park et al.~\cite{park2022192}. 
The HBM3 die photo shows that banks (including mats, BLSAs, and Sub-WL drivers) occupy about 66\% of die area.
The remaining 34\% of die area includes shared resources like through silicon vias (TSVs), IOSAs, per-bank group ECC, and column decoders.

For the tag mats, we use four smaller tag mats per data mat to reduce the row cycle time by reducing the bit line and word line lengths.
Son et al. show that the overhead of changing the aspect ratio by a factor of 4 is 19\%~\cite{son2013reducing}; however, we estimate a more pessimistic 24.3\% when we scale by 1/2 in each dimension, based on our discussions with DRAM designers.
Additionally, we only need tag mats in the even banks, reducing our overall area overhead. 
Thus, even banks require 24.3\% additional area for the tags and the data banks occupy 66\% of the die. 
So the overall impact on die size is 24.3\%~\texttimes~0.5 (only even banks)~\texttimes~0.66 (area for banks) = 8.02\%.
We also add additional area for wire routing (for example, to route Hit/Miss signals from the even bank to the odd bank), resulting in our estimate of 8.24\% die area impact.

\subsubsection{\textbf{Tag Storage Power Overhead}}

On-die tag storage and tag checking increase the power of TDRAM over standard HBM3 DRAMs.
However, \textit{TDRAM provides power benefits at the system level}, as 
(i) tags are not communicated back to the controller to perform the tag check, and 
(ii) the protocol minimizes data movement amplification, reducing the number of commands and cache line packets sent between the TDRAM and the controller.
In the HBM2 generation, 62.6\% of the power is spent moving data between the DRAM and the controller~\cite{powerSWoo}.
Avoiding data transfers associated with the tag and cache-lines, which are potentially not needed if the tag comparison result is a read miss clean for example, can be beneficial for overall memory system power.
\S\ref{sec:energy} provides an extensive power analysis of TDRAM.

\begin{table}[t]
  \centering
  \caption{TDRAM's cache operations on different accesses.}
  \label{table:operations}
  \resizebox{\columnwidth}{!}{
  \begin{tabular}{|l|l|l|l|l|}
    \hline
    \textbf{Cache Access} & \hspace{-4pt}\textbf{CMD}\hspace{-4pt} & \textbf{DQ Activity} & \textbf{HM Bus} & \textbf{Later Actions}  \\ \hline \hline

    \cline{1-1}
    Read hit to clean 
    & \multirow{6}{*}{\begin{sideways} ActRd \end{sideways}}
    & Hit Data
    & Hit 
    & None \\ \cline{3-5}

    \cline{1-1}
    Read hit to dirty 
    & 
    & Hit Data
    & Hit 
    & None  \\ \cline{3-5}
    
    \cline{1-1}
    Read to invalid 
    & 
    & None
    & Miss
    & Read main mem \& fill \\ \cline{3-5}
    
    \cline{1-1}
    Read miss to clean
    &  
    & None 
    & Miss
    & Read main mem \& fill \\ \cline{3-5}
    
    \cline{1-1}
    \vspace{-2pt}
    \multirow{2}{*}{Read miss to dirty}
    & 
    & \multirow{2}{*}{Dirty Data}
    & \multirow{2}{*}{Miss, Dirty Tag}
    & Read main mem \& fill \\ 
    
    &
    &
    &
    &  Writeback dirty data \\ \cline{3-5}

    \cline{1-2}
    Write to invalid 
    & \multirow{5}{*}{\begin{sideways} ActWr \end{sideways}}
    & Wr Data 
    & Miss
    & None \\ \cline{3-5}

    \cline{1-1}
    Write miss to clean 
    & 
    & Wr Data 
    & Miss
    & None \\ \cline{3-5}

    \cline{1-1}
    Write miss to dirty
    & 
    & Wr data
    & Miss, Dirty Tag
    & Dirty data to flush buffer \\ \cline{3-5}

    \cline{1-1}
    Write hit to clean
    & 
    & Wr Data
    & Hit
    & None \\ \cline{3-5}

    \cline{1-1}
    Write hit to dirty 
    & 
    & Wr Data 
    & Hit
    & None \\ \hline

  \end{tabular}
  }
\end{table}

\subsection{Protocol}
\label{sec:protocol}

TDRAM's command protocol is similar to traditional DRAM protocols, with modifications to minimize access latency of tags and bandwidth bloat. 
TDRAM provides new combined \textit{ActRd} and \textit{ActWr} commands that activate a row and read/write a column at both tag and data banks with an auto-precharge for close-page policy. 
These combined commands include the row and column addresses, bank group, bank, and tag address needed to determine cache hit/miss.
Internal state machines in TDRAM handle sequencing and timing of the activate and column operations to the banks and sense amps.
Read and write data appear at fixed offsets on the DQ bus from these commands, as is done in modern DRAMs.

Having a single command to access tag and data banks reduces command amplification and saves energy~\cite{micronCell,rldram3,intelRldram}.
Moreover, it simplifies the memory controller since the tag and data banks have the same number of rows and columns.
A single address is decoded for both tag and data, allowing their banks to be activated by a single command in lockstep. 
Regarding the scheduling policy of read/write requests, TDRAM's controller can adopt any policy such as first-ready first-come first-serve (FR-FCFS).
Table~\ref{table:operations} shows the operations the cache performs for each type of access.

\begin{figure}[t]
  \centering
  \includegraphics[scale=0.85]{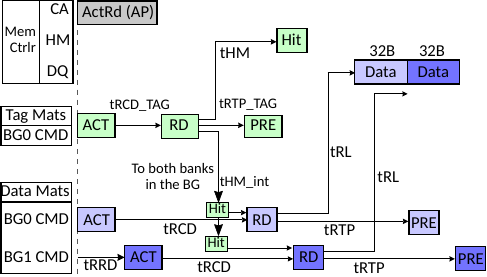}
  \caption{Timing transactions of a read operations in TDRAM. The timing
  is the same for a read miss dirty.}
  \vspace{-1em}
  \label{fig:readHit}
\end{figure}

\subsubsection{\textbf{Read Operations}}

\begin{figure}
  \centering
  \includegraphics[width=\linewidth, scale=0.8]{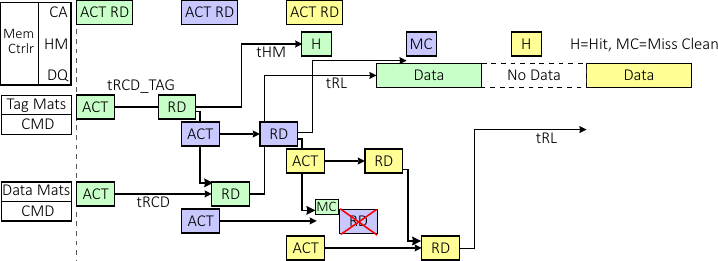}
  \caption{Timing transactions of pipelined read accesses.}
  \vspace{-1em}
  \label{fig:pipelinedReadsTrans}
\end{figure}

The low-latency tag mats allow hit-miss determination to occur before cache-line data is available.
Figure~\ref{fig:readHit} shows the timing transaction of commands involved in read operations of TDRAM.
For reads, the HM response will precede the DQ bus transfer, allowing a \textit{conditional response} based on the hit/miss result: 
(1) on a \textit{read-hit}, cache-line data is returned to the controller. 
(2) On a \textit{read-miss-clean}, no read command is issued and no cache-line data is returned to the controller. 
The unused DQ slot can be used to transfer data from the flush buffer (more details in \S\ref{sec:wrOperation}) to the controller.
(3) On a \textit{read-miss-dirty}, the dirty data is returned to the controller with the same timing on the DQ bus that would have been used to return data on a cache hit. 
The dirty tag is returned on the HM bus along with the indication that the transaction is a dirty miss. 
Figure~\ref{fig:pipelinedReadsTrans} shows the timing transactions of pipelined read accesses in different cases. 
When the controller receives miss indicator for read requests on the HM bus, it can initiate a backing store read to access the data needed (for the cache line fill and LLC response) before data (if any) arrives at the controller.
Early tag probing optimizes this further (\S\ref{sec:earlyTagProbe}).

\subsubsection{\textbf{Write Operations}}
\label{sec:wrOperation}
Writes are more complicated than reads since they must avoid overwriting a dirty cache-line with the new data on a write-miss -- a rare occurrence but one that needs to be handled correctly.
All existing DRAM caches have to serialize the cache-line data read (sending it back to the controller) and the incoming write data, for all write demands.
TDRAM avoids this serialization by implementing a flush buffer.
The flush buffer is shared among all banks, and operates similar to how a write buffer in a controller stores data to be written to the DRAMs.
The flush buffer (along with additional logic to support caching) is placed on the existing base layer which already contains logic to support HBM protocol, etc.
The base layer is not area limited, thus can support the needed buffer and logic.

\begin{figure}
  \centering
  \includegraphics[width=\linewidth, scale=0.8]{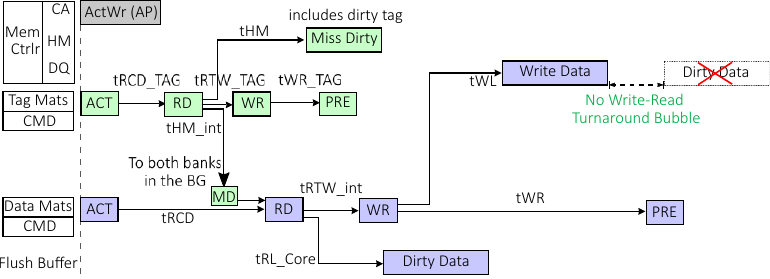}
  \caption{Timing transactions of write operations in TDRAM.}
  \vspace{-1.4em}
  \label{fig:wrDirty}
\end{figure}

TDRAM issues an ActWr command that initiates an internal tag and data access.
Once the tag comparison result arrives to the data banks, in case of hit and miss-clean, only an internal write command is issued.
If the tag check indicates a miss-dirty, an internal read command followed by an internal write command is issued.
Figure~\ref{fig:wrDirty} shows the sequence of these commands.
TDRAM places the read dirty data into the flush buffer and then writes the new data to the DRAM.
The flush buffer needs to be sized large enough such that the controller does not need to interrupt a sequence of cache writes for the sole purpose of emptying a full flush buffer, which would require insertion of a full DQ bus turnaround from write to read and then back to write direction.
When this occurs, the turnaround delays are optimized since the flush buffer is not in the DRAM core and traditional internal resource conflicts are avoided.
Since write-dirty-miss is expected to be a relatively rare event, the flush buffer can be sized modestly (16 entries or less in our simulations) to eliminate virtually any need to require a forced emptying of the flush buffer.
It is true that there still will be a small read-to-write turnaround internally to support moving the dirty data from the DRAM bank to the flush buffer, but the much larger turnaround on the DQ bus and on to the controller, can be avoided.
A sequence of cache writes from the controller would not experience any delay on the DQ bus due to the write-dirty miss.
Direct RDRAM uses a similar approach implementing a Write Buffer and a Write/Retire mechanism to minimize turnarounds due to resource conflicts in the DRAM core~\cite{hynix}.
Next we explain how TDRAM opportunistically returns the dirty data in flush buffer when DQ bus is idle or in read-state.

\paragraph*{\textbf{Unloading the Flush Buffer}}
TDRAM transmits the dirty data in the flush buffer to the controller opportunistically or on-demand, as follows: 
(i) when the DQ bus is idle, such as during \textit{refresh} operations, 
(ii) in \textit{read-miss-clean} accesses in which DQ is in read-state and is not used for data transfer, and 
(iii) if the flush buffer becomes full, the controller sends explicit read from flush buffer commands, transmitting multiple entries as a group to amortize any bus turnarounds. 
The controller has a global knowledge of the addresses in the flush buffer. 
If the DRAM cache receives a read request to any of the addresses in the flush buffer, the controller will get the data from the buffer. 
In case of a write demand to an address in the flush buffer, the incoming write demand will proceed into the DRAM cache and the controller removes the older data from the flush buffer. 
Our analysis (\S\ref{sec:fbSens}) has shown if we assume 16 entries for the buffer, transferring during read-miss-cleans and refresh cycles, prevents its overflow and any explicit read command to empty the buffer.

\subsection{Early Tag Probing Optimization}
\label{sec:earlyTagProbe}

\begin{figure}
  \centering
  \includegraphics[width=\linewidth, scale=0.8]{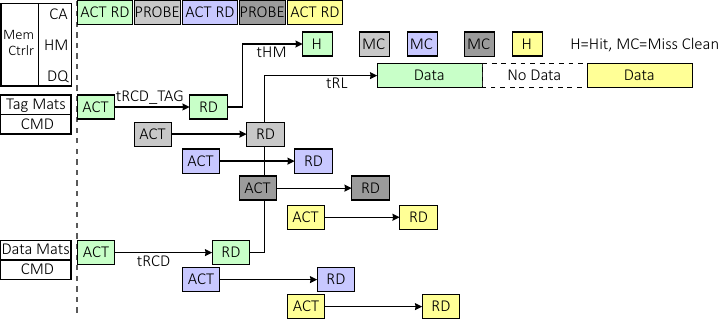}
  \caption{The timing transaction of early tag probing. The tag check results transfer from 
  tag banks to data banks is left out to make figure clear.}
  \label{fig:tagProbTrans}
  \vspace{-1.8em}
\end{figure}

The TDRAM architecture and command bus have additional unused bandwidth, because: 
(i) the timing parameters of the tag banks is shorter than the data banks; thus, the the tag banks's busy-time is less than data banks, and 
(ii) the size of the packets transferred on HM bus (few bytes) is much smaller than DQ bus (64B), while both buses work at the same frequency. 
We use this unused bandwidth for early tag probing, in which the controller can query the status of a cache-line and get an
\textit{earlier hit/miss determination} so that following actions (e.g., read main memory for read misses) can begin earlier.
Figure~\ref{fig:tagProbTrans} illustrates a set of pipelined read transactions.
While the data bus is fully occupied by back-to-back data transfers, the CA bus and HM bus are not. 
Tag probing allows a request to be sent over unused CA bus cycles to perform a tag access and comparison.

\subsubsection{\textbf{Probing Mechanism}}

Tag probing only involves accessing the tag mats and returning a result on the HM bus to the controller and does not access the cache-line data.
We refer to commands that can access both tag and data, and transferring them on the HM and DQ bus as MAIN slot commands, and tag probe requests that only access the tags and transmit status on the HM bus as PROBE slot commands.

While TDRAM without probing accelerates the tag check through fast tag bank access, the probing mechanism aims to reduce the tag check latency by \textit{minimizing the queue occupancy time of the requests waiting to be scheduled for tag access.}
For instance, if the probing indicates a miss-clean for a read demand, the request can be removed from the read queue as soon as the tag check result arrives to the controller on HM bus.
The early tag probing lowers the contention in the read buffer, requiring fewer entries and reducing the average queueing delay. 
Moreover, if a tag probe for the read requests results in a miss, the main memory access starts earlier than if the system waited for the MAIN slot for tag check.
A future MAIN slot can then be used for the cache-line fill using the data returning from the main memory.

\subsubsection{\textbf{Selection Policy}}
Once the controller finds a PROBE slot, amongst all tag check requests that can be issued at that time (i.e., no bank conflict), it picks the \textit{\textbf{youngest}} request to minimize the average queueing delay in the controller.
Even though the write packets can also use probing, TDRAM focuses on using these slots for read requests to reduce potential bank conflicts induced by early tag probing.
Our analysis has shown that the probing-induced bank conflicts are not common (less than 1\% of total demands).

\section{Evaluation Methodology}
\label{sec:method}

\subsection{Modeled System for Evaluation}
\label{sec:methodModels}
Many of the previous studies on DRAM caches~\cite{qureshi2012fundamental, jevdjic2013stacked, chou2015bear, young2018accord, jevdjic2014unison} rely on trace-based or functional-first simulators, which might not faithfully simulate the behavior of applications that take different paths depending on I/O or thread timings~\cite{eeckhout2010computer}.
In contrast, we use an execute-in-execute simulator \textit{gem5} and use full-system simulation.
Notably, prior DRAM cache research often omits full-system simulations, failing to capture OS effects.
Bin et al.'s work demonstrated that OS kernel bottlenecks can degrade memory access latency in DRAM cache systems~\cite{gao2022level}.

\textit{gem5} implements off-package memory systems via two event-driven components:
(i) memory controller responsible for receiving demands from the CPU/LLC, enqueuing them into appropriate queues, and scheduling them to access the memory device, and 
(ii) a memory interface that handles device-specific timings and operations and communicates with the memory controller~\cite{hansson2014simulating, gem5-workshop-presentation}. 
We extended \textit{gem5}'s memory system and implemented TDRAM device and DRAM cache controller~\cite{babaie2023enabling}.
This memory interface uses the DRAM timing parameters listed in Table~\ref{tab:conftable}.
We explained the timing values setup for the tag banks of TDRAM in detail in Section~\ref{sec:setTagTimings}.

We have integrated alternative DRAM cache designs into \textit{gem5} to assess the performance of TDRAM cache:
\textbf{Cascade Lake}: As the state-of-the-art commercial hardware-managed DRAM cache, we use Intel's Cascade Lake model to establish a baseline for our evaluation. 
This is a block-granule direct-mapped insert-on-miss cache storing tag and metadata in DRAM.
\textbf{Alloy}: This is a research proposal DRAM cache designed specifically to reduce hit latency~\cite{qureshi2012fundamental, chou2015bear}.
                We chose Alloy since has the most similar design principles to TDRAM.
                We do not consider predictor parts of Alloy, as they are orthogonal to our work (explained in \S\ref{sec:tagMetaStorMngmt}).
                To model Alloy's 80B burst size, we have proportionally increased the corresponding timing parameters of TDRAM (e.g., tBURST, etc.).
\textbf{TDRAM-NP}: TDRAM no probing (NP), implements all the microarchitectures we described for TDRAM in this work, except the early tag probing.
\textbf{TDRAM}: implements all the optimizations described for TDRAM, including early tag probing.
\textbf{Ideal}: We consider an ideal cache which has an architecture similar to TDRAM and knows hit/miss and metadata status in zero latency and overhead.

In order to hold a fair comparison between our approach and other DRAM caching protocols, \textit{we use the same timing parameters for modeling the DRAM device unless a parameter does not apply to a caching protocol (e.g., tRCD\_TAG in Table~\ref{tab:conftable} is only used for TDRAM cache).} 
We modeled $\frac{1}{8}$ of a target system similar to Intel's Xeon Max series~\cite{sapphire}, rounded up to 64 cores and 64~GiB of HBM (1~GiB per core, in DRAM cache mode). 
Table~\ref{tab:conftable} shows the detailed parameters of the modeled system.

\begin{table}[!h]
  \centering
  \caption{\label{tab:Config}System Configurations}
  \resizebox{\linewidth}{!}{%
    \small
    \begin{tabular}{|l|l|}
      \hline
      \multicolumn{2}{|c|}{\textbf{Processors}} \\
      \hline
      Number of cores & 8 \\
      Frequency & 5 GHz \\
      \hline
      \multicolumn{2}{|c|}{\textbf{On-chip Caches}} \\
      \hline
      Private Inst. & 32 KB \\
      Private Data & 512 KB \\
      Shared LLC & 8 MB \\
      \hline
      \multicolumn{2}{|c|}{\textbf{DRAM Cache Controller}} \\
      \hline
      Read \& Write Buffers & 64 entries each \\
      Writeback Buffer & 64 entries \\
      Controller latency & 20ns round-trip \\
      Shed. Policy & FR-FCFS \\
      \hline
      \multicolumn{2}{|c|}{\textbf{DRAM Cache (TDRAM)}} \\
      \hline
      Capacity & 8 GiB (8 channels) \\
      Peak BW & 32 GiB/s per channel \\
      Read/Write Buffer & 64 entries each \\
      \hline
      \multicolumn{2}{|c|}{\textbf{Main Memory (DDR5)}} \\
      \hline
      Capacity & 128 GiB (2 channels) \\
      Peak BW & 32 GiB/s per channel \\
      Read/Write Buffer & 64 entries each \\
      \hline
      \multicolumn{2}{|c|}{\textbf{Timing Parameters (ns) (same for all evaluated DRAM cache designs)}} \\
      \hline
      \multicolumn{2}{|p{12cm}|}{%
        Clk=2 GHz, data rate = 8Gbps, close page, RoCoRaBaCh, tBURST = 2, tRCD = 12, tRCD\_WR = 6, tCCD\_L = 2, tRP = 14, tRAS = 28, tCL = 18, tCWL = 7, tRRD = 2, tFAW = 16,  tRL\_core = 2,
        For Tag Banks in TDRAM Architecture only: tHM = 7.5, tHM\_int=2.5, tRCD\_TAG = 7.5, tRTP\_TAG = 2.5, tRRD\_TAG = 2, tWR\_TAG = 1, tRTW\_TAG = 1, tRC\_TAG = 12
      } \\
      \hline
    \end{tabular}%
  }
  \label{tab:conftable}
\end{table}

\subsection{Benchmarks}
Since we focus on large-scale computing systems for our DRAM cache design, we picked multithreaded workloads with memory footprints larger than the DRAM cache from NPB~\cite{bailey1991parallel} and GAPBS~\cite{beamer2015gap} that are used to evaluate high-performance computing systems.
Many past studies often use copies of benchmarks across multiple cores, neglecting inter-thread dependencies apparent in real-world workloads.
In contrast, we leverage multithreaded workloads to fully utilize simulated cores, enhancing realism.
We utilize the C and D class of NPB workloads and a synthetic graph for the GAPBS workloads with with 22 and 25 vertices as inputs.
Note: the performance of same workload at different classes or inputs must not be compared together, as the workload has differrent execution phases in differrent classes (\S\ref{sec:loopPoint}) that remains the same across different microarchitectures we are analyzing.
Thus, they should be seen as 28 separate workloads.

The working-set sizes of these workloads range from a few hundred megabytes to tens of gigabytes, giving different miss ratios in the 8~GiB DRAM cache.
Figure~\ref{fig:hmBreak} shows the miss ratio of these workloads.
We grouped our applications based on their miss ratios: 
(i) below 30\% miss ratio naming them low-miss-ratio, and 
(ii) above 50\% miss ratio calling them high-miss-ratio. 
There are no workloads in the middle range (i.e., no mid-miss-ratio group).

\subsection{Methodology for Experiments}
\label{sec:loopPoint}
Simulating large scale applications like NPB and GAPBS to completion is impractical, necessitating focused execution segments for each benchmark. 
Measuring work in such applications is complex due to thread interference and extended spin loops.
Traditional metrics like instruction count can cause misleading inaccurate performance measurements~\cite{alameldeen2006ipc}.
Instead, we employ a technique similar to LoopPoint, a sampling technique for multithreaded applications, tracking work progress via global loop instruction counts~\cite{sabu2022looppoint}.
LoopPoint ensures that we capture the same phases of execution on different target configurations while comparing their performance.

Following is a summary of our evaluation methodology. 
Per each workload, first, Linux kernel boots on the target system in \textit{gem5}, the program starts and continues until the start of the region of interest (ROI) of the workload using KVM CPU. 
Beginning at the ROI, we warm up the system including the CPU caches and the DRAM cache to ensure that the cold miss ratios stay a small fraction of the overall miss count (less than 1\%). 
At the end of warmup, we take a checkpoint. This process is done once per workload.
Later, we restore from the checkpoint to run all simulations using an out-of-order CPU with different DRAM cache configurations.
Using a checkpoint ensures that all runs start at the same system state for a fair comparison across different configurations.

\section{Results and Discussion}
\label{sec:Results}

\subsection{Impact of Optimizing Tag Check Mechanism}
\label{sec:tagQueDel}

\begin{figure}[t]
  \centering
  \includegraphics[width=\linewidth, scale=0.8]{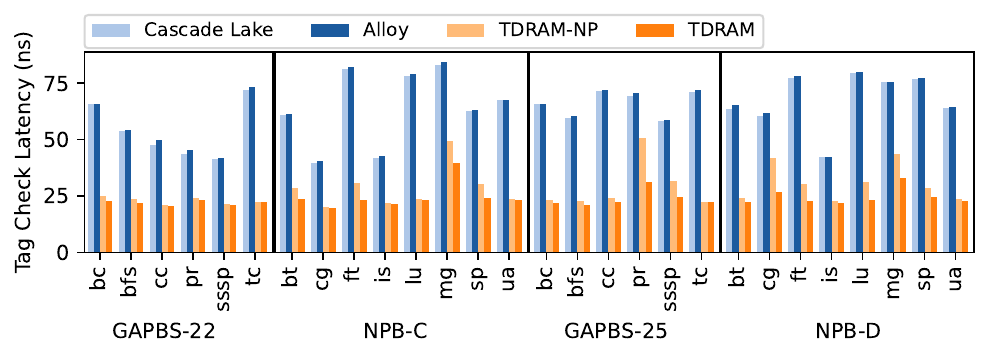}
  \vspace{-2em}
  \caption{Tag check latency comparison, the time it takes once controller receives a demand until the tag check result is ready.
           TDRAM is 2.6\texttimes~and 2.65\texttimes~faster than Cascade Lake and Alloy, respectively.}
  \label{fig:newTCL}
  \vspace{-1em}
\end{figure}

Figure~\ref{fig:newTCL} illustrates the average tag check latency for TDRAM and TDRAM-NP in comparison to Intel's Cascade Lake and Alloy caches.
Tag check latency is the time once the controller issues a tag check request to the cache until the result is ready at the controller.
The reported numbers are measured in the controller during simulation and include the queue occupancy time, DRAM cache tag access time, tag compare latency, bus latency, etc. % and the time taken to transfer the result back to the controller.
The tag access time in Cascade Lake and Alloy designs consist a read from cache line data while for TDRAM is an access to the separate tag storage of the DRAM cache.
All designs use the same timing parameters (Table~\ref{tab:conftable}) for cache line data access and TDRAM uses validated timings (based on RLDRAM) for tag storage as explained in \S\ref{sec:setTagTimings}.
In baseline design tRCD, tRL(tCL), tBURST timing parameters and for TDRAM tRCD\_TAG and tHM have the most impact in tag check latency.
Based on the simulation measurements, in both Cascade Lake and Alloy, the tag check latency falls into 40--85 ns interval, as shown in Figure~\ref{fig:newTCL}.
TDRAM-NP reduces this to 20--50 ns, and TDRAM (with early tag probing), further improves it to 19--39 ns.
TDRAM-NP achieves faster hit/miss indication across all applications compared to Cascade Lake and Alloy by parallelizing tag and data access and employing conditional data response.
TDRAM which also incorporates early tag probing, further expedites this process by opportunistically performing tag checks.
On a geomean, TDRAM's tag check is 2.6\texttimes~faster than Cascade Lake and Alloy.

Tag check latency is on the critical path of the hit and miss latencies. 
Specifically, for read demands that miss on DRAM cache, tag check latency directly impacts the LLC miss penalty, thereby affecting CPU throughput.
\textit{Improving the tag check latency accelerates the fetch of missing line from the main memory (response to the LLC), thus, reduces LLC miss penalty.} 
Figure~\ref{fig:newTCL} shows how much faster this main memory read can be issued. 

\begin{figure}
  \centering
  \includegraphics[width=\linewidth, scale=0.8]{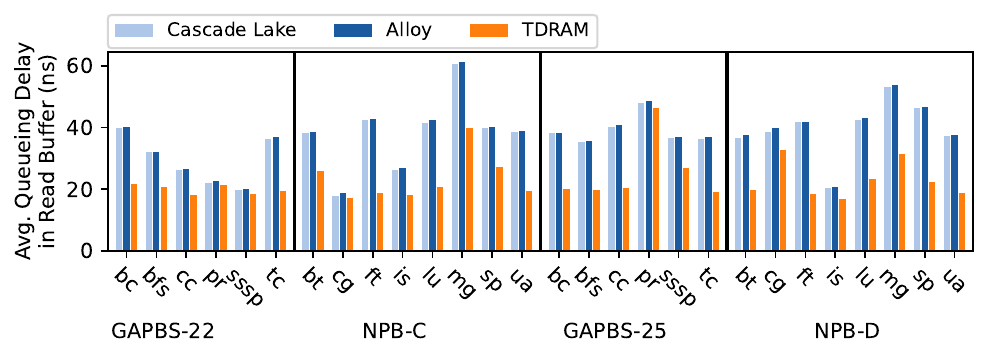}
  \vspace{-2em}
  \caption{The average queueing delay of read requests in the controller's read buffer.
           TDRAM significantly reduces the queueing delay compared to the other two designs.}
  \label{fig:queueDelTDRAM}
  \vspace{-1em}
\end{figure}

In both Cascade Lake and Alloy, for all demands, the controller issues a DRAM read for tag check, placing them in the controller's read buffer.
I.e., all read and write demands compete in the same queue for DRAM read access in their tag check process. 
This causes contention in this buffer, increasing queue occupancy time and extends the process time of read requests. 
Figure~\ref{fig:queueDelTDRAM} shows the average queueing delay of these read requests: the time taken since a read request enters the queue, until the read command for that demand is issued. 

Figure~\ref{fig:queueDelTDRAM} shows that the read queueing delay is significantly shorter in TDRAM compared to two other designs, thanks to TDRAM's early tag probing mechanism.
By employing opportunistic tag probing, TDRAM allows a read request to leave the read queue as soon as the hit-miss indicator arrives on the HM bus in the case of a miss-clean, without even activating the data bank.
This leads to fewer bank conflicts in the system and significantly impacts bank availability, resulting in reduced access time for future demands.
Note that in all read-misses in TDRAM, the controller issues a fetch from main memory immediately after seeing the miss indicator on HM bus.

The early-tag-probing benefit is totally workload dependent.
Read-misses get the most benefit from this mechanism as it accelerates the off-package memory access with 0 overhead.
If we have a smaller DRAM cache size, or workloads with larger memory footprints with the current DRAM cache size (i.e., increasing miss rates), TDRAM will benefit more from early-tag-probing.
In other words, TDRAM allows misses to happen while minimizing the miss penalty.

\begin{figure}
  \includegraphics[clip,width=\linewidth]{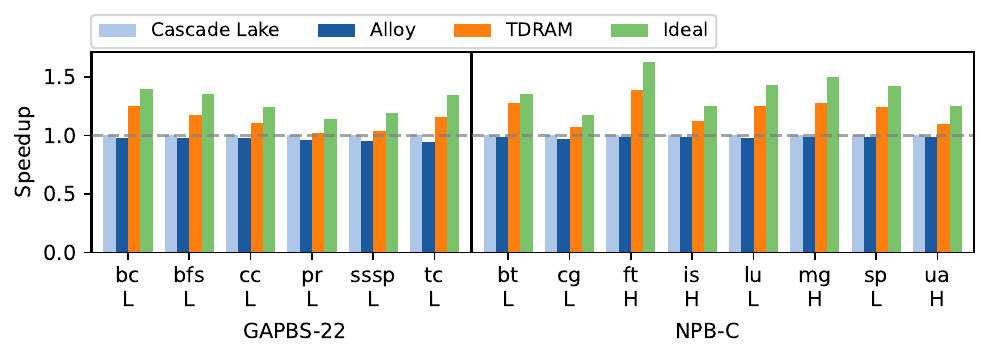}
  \vspace{-1em}
  \label{fig:perf22C}
\vspace{-0.5em} \\
  \includegraphics[clip,width=\linewidth]{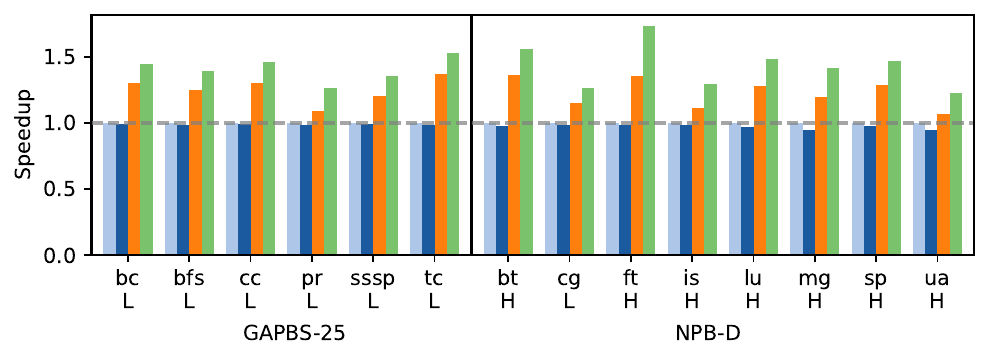}
  \vspace{-1.5em}
  \label{fig:perf25D}
\caption{System's speedup normalized to the Intel's Cascade Lake. 
TDRAM gives 1.2\texttimes~speedup on a geomean.}
\label{fig:overallPerf}
\vspace{-0.75em}
\end{figure}

\subsection{Overall Performance}
\label{sec:ovperf}

Figure~\ref{fig:overallPerf} shows the speedup of TDRAM compared to Cascade Lake and Alloy and Ideal designs.
In all workloads TDRAM outperforms Cascade Lake and Alloy, providing a geometric mean speedup of 1.20\texttimes~and 1.23\texttimes, respectively.

\begin{figure}[t]
  \includegraphics[clip,width=\linewidth]{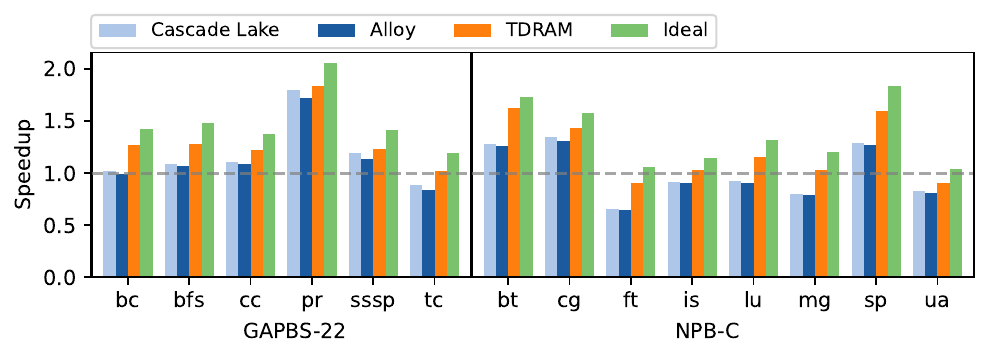}
  \vspace{-1em}
  \label{fig:perf22CNoDC}
\vspace{-0.5em} \\
  \includegraphics[clip,width=\linewidth]{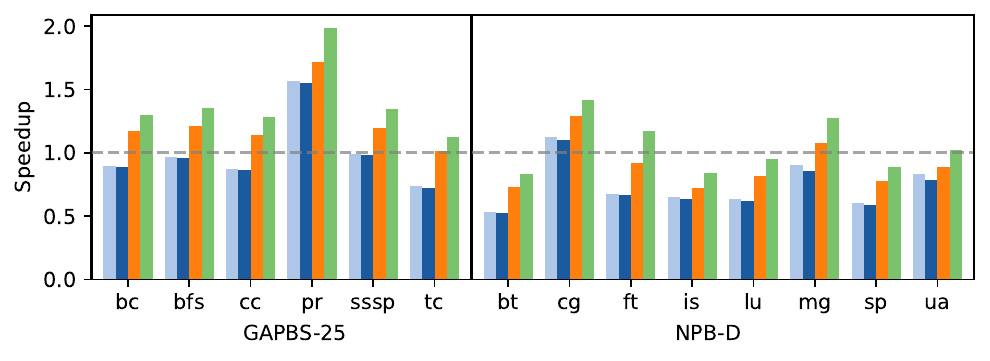}
  \vspace{-1.25em}
  \label{fig:perf25DNoDC}
\caption{Speedup of systems with DRAM cache normalized to system without DRAM cache.
  TDRAM provides 1.11\texttimes~speedup, while Cascade Lake and Alloy cause 8\% and 10\% slowdown, respectively.}
\label{fig:overallPerfNoDC}
\vspace{-1.2em}
\end{figure}

As discussed in \S\ref{sec:tagQueDel}, TDRAM effectively reduces tag check latency and read demands queueing delay.
This improvement positively impacts the hit and miss latency of the DRAM cache and the miss penalty of LLC.
Consequently, the overall performance of the system is enhanced compared to existing designs, as evident in Figure~\ref{fig:overallPerf}.
The ideal cache offers tag check results with zero latency, eliminating the need to endure queueing delay and DRAM access latency for tag checks.
In essence, the ideal cache sets a performance upper bound for caching, and Figure~\ref{fig:overallPerf} demonstrates that TDRAM closely approaches this ideal.

Figure~\ref{fig:overallPerfNoDC} compares the speedup of the aformentioned four systems to the same system that has only a main memory (no DRAM cache).
As the figure shows for applications with lower miss ratios DRAM caching improves system overall throughput.
This improvment decreases as the miss ratio increases due to the miss penalty of DRAM cache that involves main memory acccess.
Analyzing the data, Intel's Cascade Lake and Alloy caches cause a geomean \textit{slowdown} of 8\% and 10\%, respectively.
In contrast, TDRAM provides an overall \textit{speedup} of 1.11\texttimes, primarily due to optimizations in hit latency and miss penalty within its protocol.

\subsection{TDRAM's Energy Improvement}
\label{sec:energy}
Prior work has defined bandwidth bloat factor as: total number of bytes moved, over total useful bytes moved~\cite{chou2015bear}.
Table~\ref{table:bwbloat} shows the bandwidth bloat factor for the evaluated designs.
Our results shows that TDRAM reduces the bandwidth bloat factor upto 39.9\% and 25.1\% compared to Alloy and Cascade Lake, respectively.

\begin{table}[h]
 \centering
 \caption{Bandwidth Bloat Factor}
 \label{table:bwbloat}
 \begin{tabular}{|p{2.3cm}|p{2cm}|p{2cm}|}
  \hline
  \textbf{Design} & \textbf{Low-Miss Ratio} & \textbf{High-Miss Ratio} \\
  \hline
  \hline
  Alloy & 1.68 & 3.43 \\ \hline
  Cascade Lake & 1.35 & 2.75 \\ \hline
  TDRAM & 1.13 & 2.06 \\
  \hline
      \multicolumn{3}{|c|}{\textbf{TDRAM Reductions}} \\
  \hline
  Over Alloy & 32.7\% & 39.9\% \\ \hline
  Over Cascade Lake & 16.3\% & 25.1\% \\ \hline
 \end{tabular}
\end{table}

\begin{figure}[t]
  \includegraphics[clip,width=\linewidth]{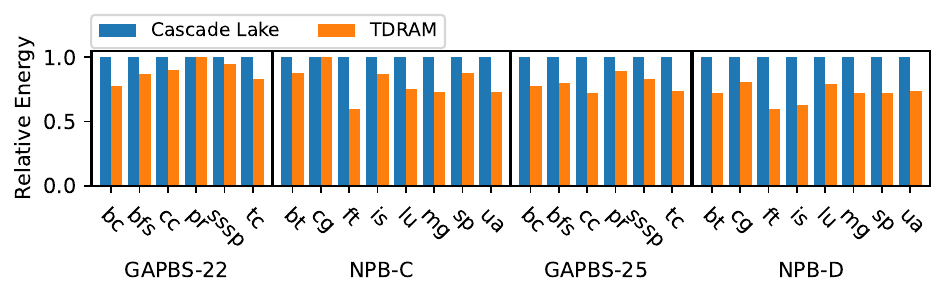}
  \vspace{-2em}
  \caption{Relative energy consumption of TDRAM compared to baseline. TDRAM reduces 
  energy by 21\% on average compared to the baseline.
}
\label{fig:relEnergy}
\vspace{-1em}
\end{figure}

As the bandwidth bloat factor increases, more energy is consumed since more data is transferred. 
TDRAM eliminates unnecessary data transfers, saving energy by reducing the total data movement, while servicing the same number of memory demands as the other two designs. 
To analyze the energy consumption of TDRAM, we developed an HBM3 power model using HBM2 power data in~\cite{o2017fine} and scaled it for HBM3 speeds and timings (Table~\ref{tab:conftable}).
Processor interface power is calculated from our validated HBM3 PHY design.
Compared to a standard HBM3 DRAM, TDRAM’s power is increased to account for on-die tag storage and associated operations, and both DRAM cache and processor interface power are increased for the additional signals and HM buses and associated logic.
Figure~\ref{fig:relEnergy} shows the relative energy consumption (power \texttimes benchmark runtime) of TDRAM and Cascade Lake DRAM cache.
We are not showing Alloy's energy consumption as it is much higher than Cascade Lake.
On average, TDRAM saves 21\% energy compared to the Cascade Lake due to reducing bandwidth bloat. 
Applications with high number of write-hits or read/write miss-cleans demands, such as $ft$ and $is$, show more energy savings with TDRAM's.

\begin{figure}[t]
  \includegraphics[clip,width=\linewidth]{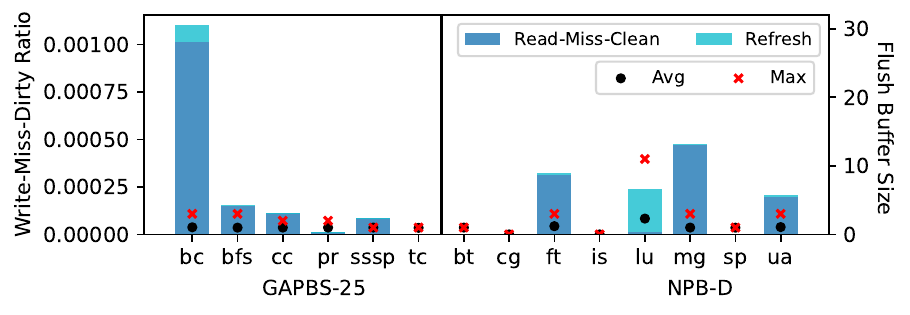}
  \vspace{-2em}
  \caption{Flush buffer size sensitivity test.
  The left axis is the ratio of write-miss-dirty demands out of total requests that DRAM cache received.
  Each bar is broken into different colors depicting the ratio of dirty data in the flush buffer that was unloaded during read-miss-cleans or refresh cycles.
  The right axis shows the maximum and average occupancy of flush buffer when it had 32 entries total.
}
\label{fig:flushBufferSens}
\vspace{-1.5em}
\end{figure}

\subsection{Flush Buffer Size Sensitivity Analysis}
\label{sec:fbSens}

We assessed the sensitivity to the flush buffer size with 8, 16, 32, and 64 entries.
The results indicate that the flush buffer consistently avoids becoming full, preventing TDRAM stalls to empty the buffer, except for a minor exception with $lu$ in NPB-D with a buffer size of 8.
In this case, TDRAM stalled only 13 times, resulting in negligible performance overhead.
Figure~\ref{fig:flushBufferSens} shows the total write-miss-dirty accesses ratio in NPB-D (which has the largest memory footprint and highest miss rate among all workloads we tested) with a flush buffer size of 32.
The figure represents how the dirty data was transferred from the flush buffer to the controller, via read-miss-cleans, refresh cycles, or explicit command to read from flush buffer (which is zero).

Write-miss-dirty accesses in the DRAM cache should be rare, since write-misses in general are the LLC write-backs, that previously required a read in the DRAM cache.
This expectation is confirmed by our results, as shown in the left Y-axis the write-miss-dirty ratios are extremly low.
The results also shows that most applications heavily rely on read-miss-clean accesses to unload the flush buffer.
Notably, $lu$ and $bc$ efficiently use refresh cycles to unload the flush buffer.
This data confirms the effectiveness of TDRAM's opportunistic behavior in unloading the flush buffer to minimize data transfer overhead.
The flush buffer's average occupancy is 5, with a maximum of 12.
Setting the buffer size to 16 prevents TDRAM stalls.
Thus, the overhead of flush buffer is minimal.

\subsection{Link Latency Case Study}
\begin{figure}
  \includegraphics[clip,width=\linewidth]{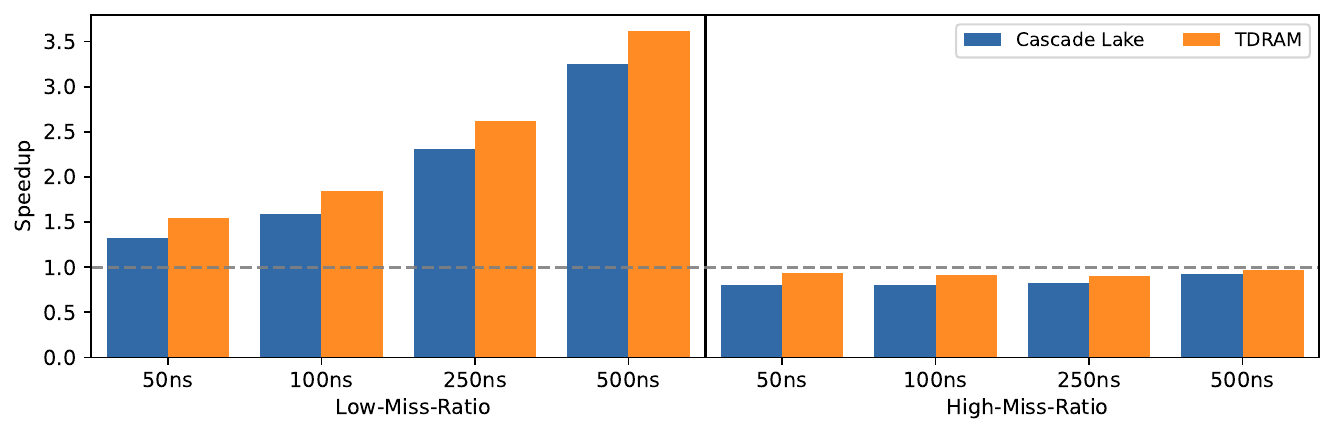}
  \vspace{-1.9em}
  \caption{Speedup of systems with DRAM cache in disaggregated systems normalized to the same system having a remote main memory only (no DRAM cache).}
\label{fig:linkLatAll}
\vspace{-1.5em}
\end{figure}

We assume incorporating DRAM caches in disaggregated memory systems where it accesses the main memory through an interconnect such as CXL.
We consider round-trip link latencies of 50~ns, 100~ns, 250~ns and 500~ns. 
Figure~\ref{fig:linkLatAll} shows the speedup of the DRAM cache systems compared to the same system having a remote main memory only (no DRAM cache) and the link latency of this main memory also varies from 50--500~ns.

As Figure~\ref{fig:linkLatAll} shows, in all cases TDRAM outperforms Cascade Lake DRAM cache.
We are not showing the Alloy cache data as it was worse performant than Cascade Lake in all cases.
For low miss ratio applications, Cascade Lake has an overall speedup of 1.99\texttimes~while TDRAM increases this to 2.27\texttimes, based on a geometric mean of all speedups.
For high miss ratio applications, the speedup is challenging due to the extended miss penalty that includes the interconnet's latency. 
Cascade Lake DRAM cache system causes 17\% slowdown; however, TDRAM reduces it to 7\%.

\subsection{Set-Associative TDRAM}
TDRAM's on-die tag comparison is beneficial for set-associative caches since it avoids transferring multiple tags in a set to the controller for comparison. 
In general, set-associativity is helpful for applications with high number of miss conflicts.
However, our analysis showed the HPC workloads we tested barely have miss conflicts on DRAM cache.
Thus, they did not get significant performance improvement from set-associativity compared to direct-mapped TDRAM.
Our results showed both direct-mapped and 16-way set-associative TDRAMs have similar speedup (over the same system having a main memory only) for tested workloads.

\vspace{-1ex}
\section{Related Work}
\label{sec:related}
Table~\ref{tab:related} and \S\ref{sec:tagMetaStorMngmt} provides a comparison of TDRAM with prior work.
Loh and Hill~\cite{loh2012supporting} proposed one of the earliest block-based DRAM cache where tag and data access were stored in the same row with a MissMap to avoid accessing the DRAM cache on predicted misses.
Alloy~\cite{qureshi2012fundamental} reduces latency by streaming data and tags together in a single burst.
Furthermore, they introduced a memory access predictor, which incurred less overhead compared to the MissMap technique.
Retagger~\cite{bojnordi2019retagger} uses tags in the controller to mitigate the DRAM row buffer miss cost.
RedCache~\cite{behnam2022adaptively} adapts at runtime to start and stop caching for individual blocks.
While these works have explored different approaches for storing tags and data in DRAM caches, they all require the tags to be moved to the controller for tag comparison and checks to be performed.
R-Cache proposed to use RRAM memory for on-die tag storage~\cite{behnam2018r}.
Due to longer latency of RRAM compared to DRAM, it can extend the tag check latency, exacerbating the hit and miss latencies of DRAM cache.
In contrast, our work modifies the DRAM microarchitecture to enable tag checks to be performed inside the DRAM, thereby reducing the data movement overhead and improving overall cache efficiency.

The Footprint Cache~\cite{jevdjic2013stacked} and 
Unison Cache~\cite{jevdjic2014unison} blend block and page-based designs in a hybrid architecture.
This lowers off-chip traffic compared to page-based designs, with high hits, low latency, and minimal tag overhead.
Coarse-grain tracking leads to bandwidth waste and poor utilization of cache capacity in contrast to block-based caches like TDRAM.

Several works~\cite{yu2017banshee,kotra2018chameleon,kim2023nomad} combine software and hardware techniques for DRAM caching.
Also, Hong et al. proposed a DRAM cache specifically for GPUs working with storage-class memories\cite{hong2024bandwidth}.
We envision potential software/OS integration benefits for TDRAM, as well as GPU-specific changes which are directions we plan to explore.

To our knowledge, the only work to leverage HBM's embedded logic die for cache management is Stockdale et al.'s~\cite{stocksdale2017architecting}.
They enhance the base HBM DRAM layer, introducing a cache result signal and reserving one pseudo-channel for tags.
The eTag DRAM cache uses eDRAM storage on the processor die, with tag comparison preceding DRAM cache access~\cite{yang2016etag}, but eTag cannot scale with increased off-chip capacity as eDRAM size limits the data cache capacity.
Hameed et al. proposed a DRAM cache with a separate tag and data storage that relies on a predictor and a Data-Absence-Table~\cite{hameed2020improving}.
These speculation-based designs are orthogonal to TDRAM.
TDRAM minimizes amplification, using the HM bus for cache outcomes and tag transfers, and employs tag probing to mitigate read miss impact.
TDRAM puts tags and data on each channel which scales with cache capacity.

\vspace{0.5ex}
\section*{Conclusion}
\vspace{0.5ex}
In this paper we introduced TDRAM, tag-enhanced energy-efficient DRAM for caching, to optimize caches hit and miss latencies. 
We showed TDRAM's 1.2\texttimes speedup and 21\% energy saving over commercial designs (Intel's Cascade Lake) and research proposal (Alloy). 
TDRAM can bridge performance gaps between LLC DRAMs and remote memories in heterogenous/disaggregated systems.

%%%%%%% -- PAPER CONTENT ENDS -- %%%%%%%%

%%%%%%%%% -- BIB STYLE AND FILE -- %%%%%%%%
\bibliographystyle{IEEEtranS}
\bibliography{refs}
%%%%%%%%%%%%%%%%%%%%%%%%%%%%%%%%%%%%

\end{document}